\documentclass[12pt,preprint]{aastex}

\begin{document}

\title{Galaxy interactions in the Hickson Compact Group 88}

\author{Noah Brosch\altaffilmark{}}
\affil{The Wise Observatory and the Raymond and  Beverly Sackler School of Physics and
Astronomy, the Faculty of Exact
Sciences, \\ Tel Aviv University, Tel Aviv 69978, Israel}
\email{noah@wise.tau.ac.il}


\begin{abstract}
I present observations of the Hickson Compact Group 88 (HCG88) obtained during the commissioning of a new 28-inch telescope at the Wise Observatory. This galaxy group was advertised to be non-interacting, or to be in a very early interaction stage, but this is not the case.  The observations reported here were done using a ``luminance'' filter, essentially a very broad R filter, reaching a low surface brightness level of $\approx$ 26 mag arcsec$^{-2}$. Additional observations were obtained in a narrow spectral band approximately centered on the rest-frame H$\alpha$ line from the group. Contrary to previous studies, my observations show that at least two of the major galaxies have had significant interactions in the past, although probably not between themselves. I report the discovery of a faint extended tail emerging from the brightest of the group galaxies, severe isophote twisting and possible outer shells around another galaxy, and map the HII regions in all the galaxies.
\end{abstract}



\section{Introduction}
Hickson (1982) identified 100 compact groups of galaxies  by searching the Palomar Observatory Sky Survey (POSS) red prints. The selection criteria he used combine the number of members of comparable luminosity ($\geq 4$), the degree of isolation of the group, and the total magnitude of these galaxies averaged over the area they cover. The last two criteria are designed to exclude clusters of galaxies, while selecting high-galaxy-density regions. Since the projected galaxy density in a Hickson Compact Group (HCG) can reach up to 10$^8 h^{-2}$ gal Mpc$^{-2}$ (Hickson et al. 1992), the HCGs generated significant interest and numerous papers since their recognition, while some of the groups were found not to be real based on redshifts, being just chance superpositions. The ``real'' groups offer a chance to study the formation and evolution of galaxies in a very dense environment that is not a galaxy cluster. Moreover, as Amram et al. (2002) pointed out, the HCGs might be the nearby-Universe analogs of environments in the young Universe. Ponman et al. (1996) mentioned that, since the galaxy densities in HCGs are very high, the galaxies within one such group should merge ``within a few crossing times''.

Hickson's compact group 88 (HCG88) included originally four members, but following the redshift survey of some HCGs by de Carvalho, Ribeiro, Capellato and Zeff (1997) two more (fainter) members were added (Ribiero et al. 1998). HCG88 was also identified as a compact group based on a 2MASS survey by D{\'{\i}}az-Gim{\'e}nez et al. (2012). At the average recession velocity of 6126 km s$^{-1}$ the group is at a distance of about 84 Mpc (h=0.73) and the scale is 24.6 kpc arcmin$^{-1}$.

The six known members, and more distant but relatively nearby galaxies possibly associated with the group, found from a NED\footnote{http://ned.ipac.caltech.edu/; search performed in July 2015} search out to a groupocentric distance of 60 arcmin$\simeq$1.5 Mpc and within 500 km s$^{-1}$ of the catalog group redshift, are listed in Table~\ref{t.group}. The table shows the six previously-identified group members and the two other much more distant galaxies, and indicates that these two have the right recession velocity to be considered group outliers.
Extending this search even further, to 3$^{\circ}$ from HCG88, while keeping the redshift difference within 500 km s$^{-1}$ of the average group redshift, adds tens of objects, all possibly connected with the galaxy cluster UGCl 458 at the edge of the searched region, almost 5 Mpc away in projected distance. UGCl 458 is a poorly-studied cluster with more than 550 members at a NED redshift of 5925 km s$^{-1}$

\begin{table}
\vspace{0.5cm}
\label{t.group}
\begin{footnotesize}
\begin{center}
\begin{tabular}{|c|c|c|c|c|c|c|c|}
\hline
Galaxy & RA & Dec. & v$_{\odot}$ & Type & D & SDSS-r & Remarks \\
 & (J2000) & (J2000) & (km s$^{-1}$) & & (') &  & \\
\hline
N6978 & 20:52:35.4 & -05:42:40 & 6033 & Sb (a) & 4.2 & 13.00 &  HCG88A \\
N6977 & 20:52:29.7 & -05:44:46 & 6192 & SBb (a) & 1.9 & 13.23 & HCG88B \\
N6975 & 20:52:26.0 & -05:46:20 & 5956 & SAB(r)bc? (b)  & 1.2 & 14.07 & HCG88C \\
MCG-01-53-014 & 20:52:12.8 & -05:47:54  & 6041 & S? (b), Sc (e) & 3.5 & 14.96 & HCG88D \\
SDSS J205223.18-054325.7 & 20:52:23.2 & -05:43:26 & 5805 & dwarf? (d) &  2.0 &  17.33 & \\ 
SDSS J205224.43-054714.2 & 20:52:24.4 & -05:47:14 & 6091 & dwarf? (c) & 1.8 & 17.46 & \\ 
2MFGC 15822 & 20:54:12.2 & -05:11:32 & 6163 & Scd(f) (c) &  43.5 & 16.36 & \\
2MASX J20493219-0626432 &  20:49:32.2 & -06:26:43 & 5979 & dwarf (d) &  59.2 & 16.11 & \\
\hline
\end{tabular}
\end{center}
Notes to Table~\ref{t.group}: D is the distance of each object from the nominal center of the HCG88 group. The morphological classifications are from (a)  Coziol et al. (1998); (b) RC3; (c) NED/DSS1; (d) own estimation; (e) Plana et al. (2003). The SDSS magnitudes are in the $r$ band, Petrosian, in the AB system. The last two objects are distant but have similar redshifts to the objects in the group.
\end{footnotesize}
\end{table}

Table~\ref{t.group} and the group image shown in Figure~\ref{fig:HCG88} demonstrate that this is an elongated collection of late-type galaxies. In particular, SDSS J205224.43-054714.2 is probably a late-type dwarf since it appears diffuse on my images and its SDSS image is bluish and elongated while its SDSS spectrum is blue showing emission lines. SDSS J205223.18-054325.7 is also a dwarf galaxy, but it shows signs of an earlier stellar population with an SDSS absorption spectrum. The projected linear distance between HCG88A and HCG88D is $\sim$190 kpc; assuming that the group has s surface 200 kpc long and 50 kpc wide, the projected surface density of these four major components of HCG88 is therefore $\sim$400 galaxies Mpc$^{-2}$, higher than the value of 160 galaxies Mpc$^{-2}$ listed for this group in Verdes-Montenegro et al. (2001). 

Given its linear structure and low radial velocity dispersion,  HCG 88 has been claimed to be a filamentary structure (Hernquist, Katz \& Weinberg 1995; Da Rocha \& Mendes de Olivera 2005);  similar galaxy filamentary structures were identified by Zitrin \& Brosch (2008) and by McQuinn et al. (2014). 
Filamentary structures in the three-dimensional distribution of galaxies might be related to the ``cosmic web'' of dark matter onto which intergalactic matter collects and forms galaxies. A Lyman-$\alpha$ filament was revealed at z$\simeq$2.3 (Cantalupo et al. 2014), but filaments such as those mentioned above, which are relatively nearby, might be even more useful since faint galaxies would be detectable there with reasonably-sized telescopes.
The galaxy filaments identified by Zitrin \& Brosch and by McQuinn et al. are very nearby, at a recession velocity of a few 100 km s$^{-1}$. The possible filament formed by HCG 88 and discussed here is more distant, at about 6000 km s$^{-1}$, thus its study would be limited to the brighter members only.

Verdes-Montenegro et al. (2001) studied some compact groups with VLA HI mapping and proposed that the evolution of HCGs starts with objects similar to HCG88 (their Phase I) where most of the HI is in the disks and only a small fraction, 2.9$\times10^8$ M$_{\odot}$, is in the tidal tails. Since they found that most of the HI is located in the galaxies' disks, with only a small fraction in the tidal features, they concluded that HGC88 should be considered to be at a very early stage of interaction. Da Rocha \& Mendes-de Olivera (2005) imaged the main galaxies in HCG88 with the CFHT (FOV=2'4$\times$3'.1) and used wavelet-decomposition techniques in an attempts to detect diffuse intergalactic light. They put an upper limit to such light at $\mu_B$=29.1 mag arcsec$^{-1}$ and $\mu_R$=27.9 mag arcsec$^{-1}$, from which they reached a similar conclusion to that of Verdes-Montenegro et al. (2001),  that the group is only at a very early stage of interaction.

The ionized gas kinematics in HCG88 were studied by Plana et al. (2003) based on Fabry-P\'{e}rot observations with the ESO and CFHT 3.6 telescopes. They found a clumpy medium in HCG88A, B and D, and a possible three-armed structure in HCG88C, while not reporting ``signs of previous or current interactions''. Based on the kinematic structures (see their Table 5) Plana et al. argue that HGC88C, and possibly also HCG88D, might be undisturbed, whereas HCG88A may have suffered a mild interaction, with the entire group showing no merger indication. They concluded that HCG88 ``may be an unevolved group'' but list in their Table 5 the high IR luminosity and central activity of N6978 as indicators of interaction. For HCG88B they list five such interaction indicators, including also the highly disturbed velocity field, the changing position angle along the major axis, and the misalignment of gas and stars along the major axis.

These claims that HCG88 may be unevolved or only slightly evolved, whether the galaxies are interacting or not, argue in favour of looking more carefully at this group in the context of the ``nature vs. nurture'' problem in galaxy evolution. Specifically, it would be useful to identify supporting evidence that this group is indeed unevolved or, alternatively, disprove previous claims by identifying signs of interaction. One possible way to achieve this is by using images that cover the entire group and significant parts of its surroundings, while reaching low surface brightness (LSB) levels. This technique already revealed interesting LSB features around other galaxies (e.g., near N4449 by Rich et al. 2012, or near M51 by Watkins, Mihos, \& Harding 2015).

Deep imaging, beyond what is currently available in public data bases such as SDSS, has the potential of revealing signs of interaction such as LSB stellar streams, ``heated'' galaxy disks and extended galaxy halos (e.g., Cooper et al. 2010; Purcell, Kazantzidis \& Bullock 2009, Purcell, Bullock \& Kazantzidis 2010). Such searches have been restricted mainly to resolved-star studies of the Milky Way and M31 (e.g., the Pan-Andromeda Archeological Survey=PandAS; Ibata et al. 2013, Martin et al. 2013), since extending them to more distant objects requires significant detection capability for very LSB extended features. If  $\Lambda$CDM is correct, then dark matter (DM) halos of galaxy-size should be constructed from the steady accretion of smaller, dwarf-galaxy sized bits of DM and regular matter (e.g., Fakhouri, Ma \& Boylan-Kolchin 2010,  based on the Millenium simulation). Traces of such accretion events should be detectable, if deep imaging is available.

The structure of this paper is as follows. Section~\ref{sec.obs} describes the observations and their reductions. Section~\ref{sec.proc} details the reduction process, which revealed faint structures near some of the galaxies and mapped the distribution of HII regions. The results are detailed in the same section, and in Section~\ref{sec.discuss} HCG88 is discussed in the context of published scenarios for its formation and evolution. Section~\ref{sec.summary} summarizes this paper.

\section {Observations and data reduction}
\label{sec.obs}

HCG88 was observed during the commissioning of a new telescope at the Wise Observatory, Israel. The telescope, called the ``Jay Baum Rich telescope'' (hereinafter JBRT), is extensively described in Brosch et al. (2015). Briefly, it is a 28-inch (0.7-m) Centurion-28 prime-focus f/3.1 reflector, imaging a $\sim$one degree wide field of view onto a CCD camera behind a doublet field-corrector lens. The camera is a Finger Lakes Instruments ProLine 16801 equipped with a five-position filter wheel and thermoelectrically cooled to approximately --30C using water assist. The plate scale is 0".83 pixel$^{-1}$, the images are digitized to 16 bits, the readout is done at 8MHz, and the dark counts are $\sim0.05$ sec$^{-1}$ pixel$^{-1}$. The 4k$\times$4k chip covers a bit less than one square degree of the sky and has a peak quantum efficiency of 67\% at 661 nm.

\begin{figure}[h]
\centering{
  \includegraphics[width=7.5cm]{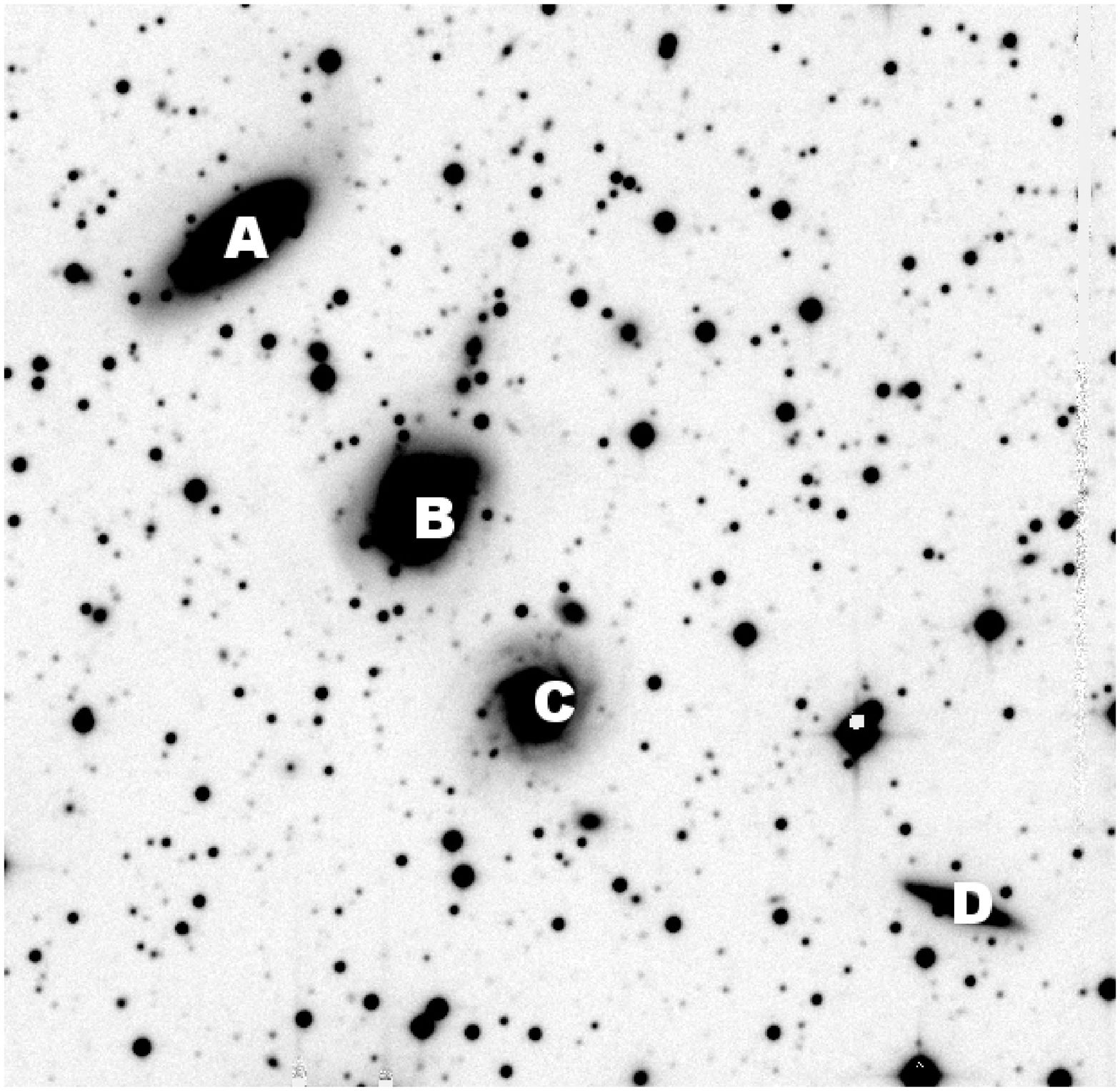}
  \includegraphics[width=8.2cm]{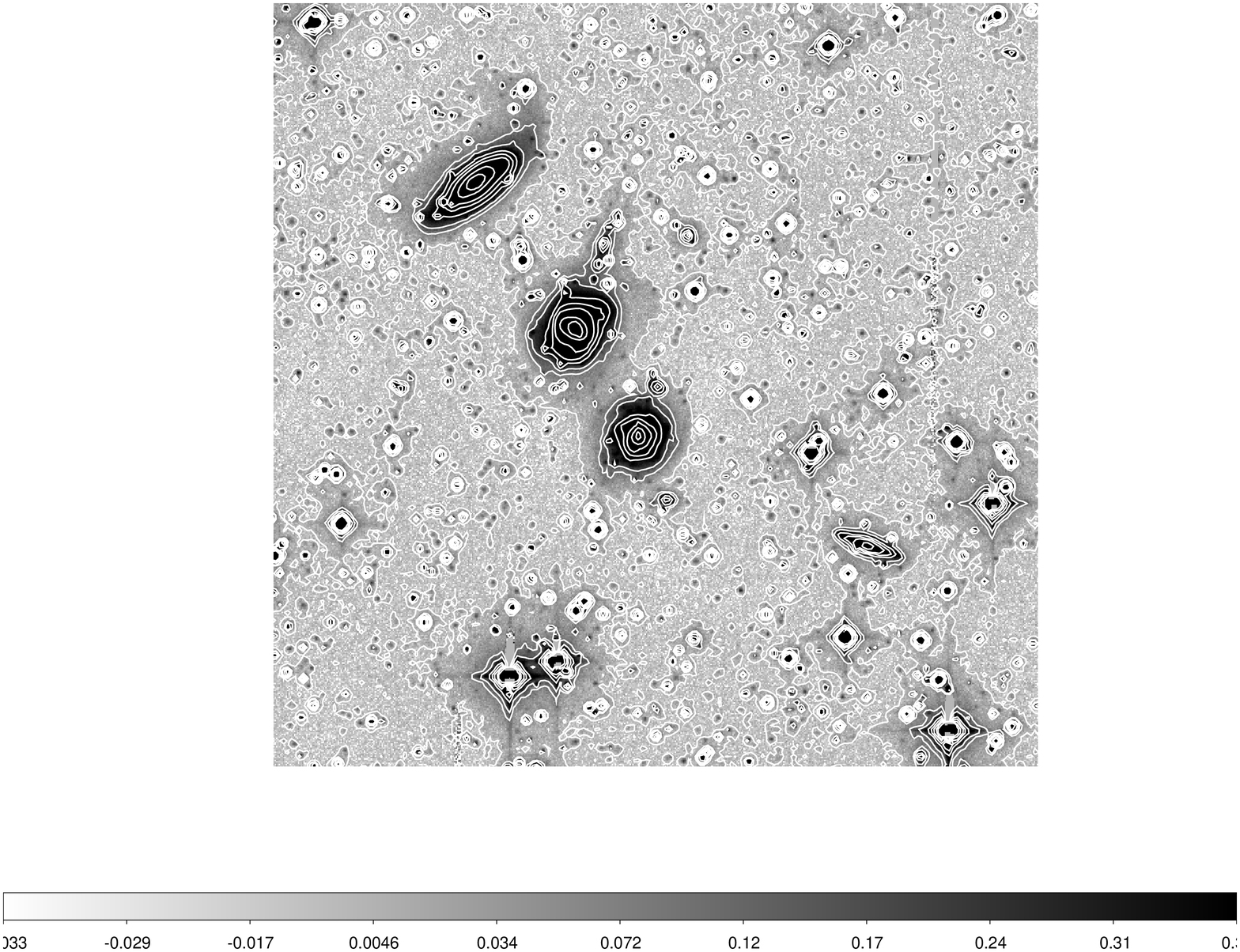}
  \caption{{\it Left panel:} The Hickson Compact Group 88. This is a co-added composite of 37 images, each exposed for 300-sec through the L-band filter, unsharply-masked and contrast-stretched, to reveal low-surface brightness details in the galaxies. North is up and West is to the right; the four main group components A, B, C and D are aligned in this order from top-left to bottom-right of the image and are marked by labels. {\it Right panel:} Contours of the image of the entire group emphasize the faint northward extension from N6978 (HCG88A), as well as a straight projection from the $\sim$face-on disk of N6977 (HCG88B) that is, however, a superposition of background galaxies. Note the extreme change of the position angle of the major axis of this galaxy when going to fainter surface brightness levels.}
  \label{fig:HCG88}}
\end{figure}

The observations were performed during the early commissioning stage of the telescope on June 26, 2014 when HCG88 was followed from $\sim$3$^h$ East of the meridian to about the same distance West of the meridian. The observations were successive 300-sec exposures which were dithered in arbitrary directions by at most 20 pixels. 
 A ``luminance'' filter, hereafter called L-filter (Baader Planetarium UV/IR-cut), was used throughout. This filter selects a $\sim$250-nm wide spectral region centered on $\sim$560 nm and with in-band transmission better than 94\%. With its sharp cutoffs at $\sim$400 and $\sim$690 nm, it reduces the sky background by excluding the blue-UV light originating from Hg street lamps as well as the red atmospheric emission. A total of 37 L-band images were obtained, were processed, and were combined into a final image shown in the left panel of Fig.~\ref{fig:HCG88}. The reduction process used the THELIGUI package (Erben et al. 2005; Schirmer 2013) and included bias and dark subtraction, flat fielding using morning twilight exposures, and astrometric registration.

\begin{figure}[h]
\centering{
  \includegraphics[width=15cm]{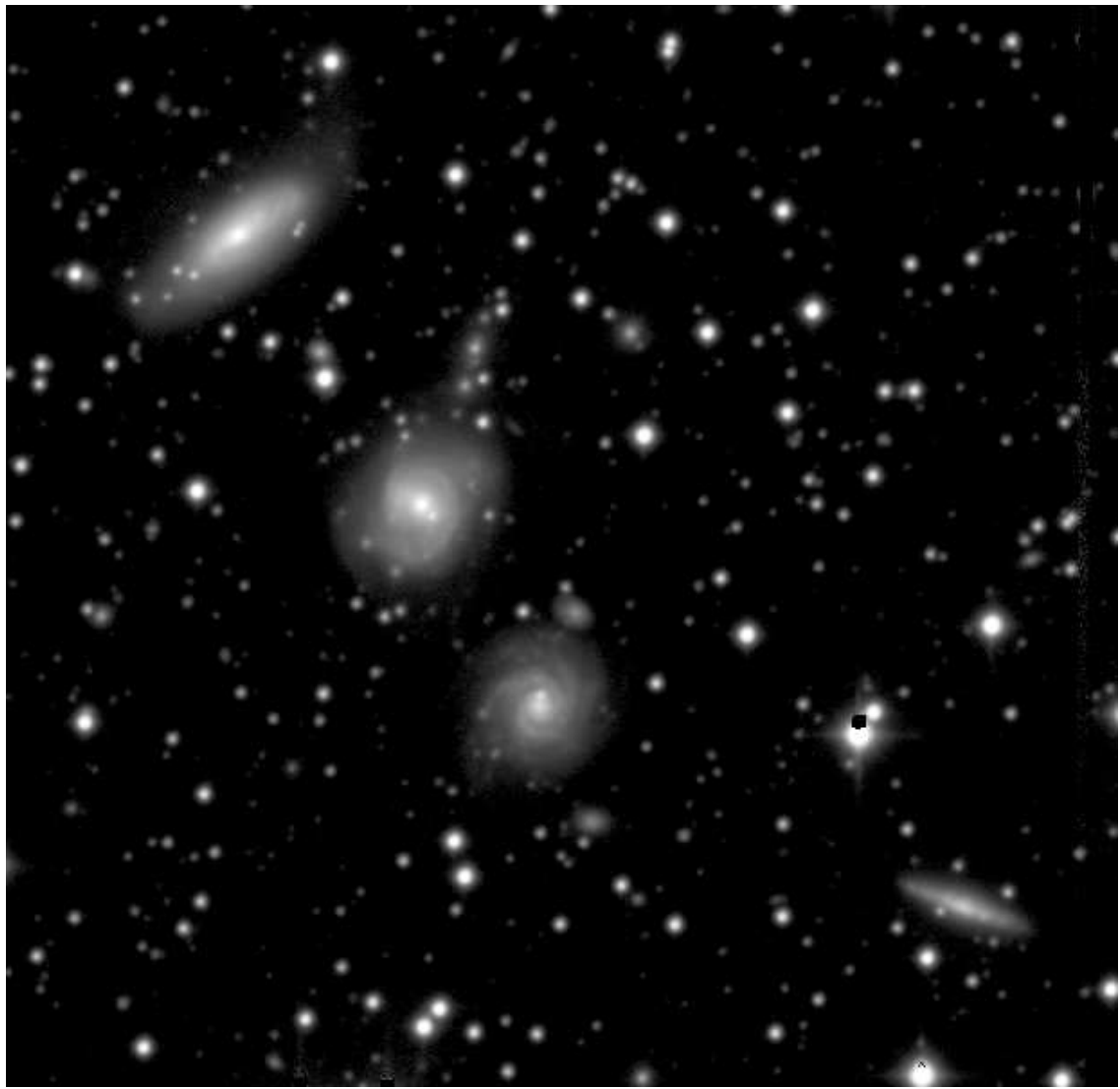}
  \caption{HCG88 displayed as a positive image and with a strong contrast stretch to show internal details of the galaxies.}
  \label{fig:HCG88_pos}}
\end{figure}

Figure~\ref{fig:HCG88_pos} shows a positive, stretched-contrast version of Figure~\ref{fig:HCG88}. Note, in particular, the sharp edges of the South-East part of HCG88B, in contrast with the fuzzy edges of the three other galaxies and the elongation of its faint outer contours $\sim$perpendicular to its inner, brighter regions.

In addition, rest-frame H$\alpha$ images were collected using the WO 40-inch telescope equipped with a Princeton Instruments CCD. Eight images, each exposed for 900 sec, were collected on 13 October 2014 along with bias, dark exposures and twilight flat fields. The images were exposed through the WO H$\alpha$-5 filter with a central wavelength of 6696\AA\, and a full-width at half-maximum of 58\AA\,, excellently matching the redshift of HCG88. At the peak of the profile, this filter's transmission is 62\%. These images were processed with THELIGUI just as the L-band images.

The H$\alpha$ and the L-band images were put on the same plate scale, were registered and were sky-subtracted. The L-band image was scaled so that the stars would have approximately the same total counts in both images. The scaled L-band image was subtracted from the H$\alpha$ image leaving only the line emission at the redshift of the object. Note that this method of compensating for the continuum emission, using a broad-band filter, had been studied by Spector, Finkelman \& Brosch (2012) and was shown to produce sometimes measurement errors by as much as 40\% in the line equivalent width if the colors of the comparison stars used to scale the continuum are not properly compensated. However, this is not the case for low-redshift objects such as HCG88 where the errors in subtracting without compensating for the colors of the stars could induce errors of a few percent only.

\section {Analysis and results}
\label{sec.proc}

Inspecting the deep L-band image of the group (see the left panel of
 Fig.~\ref{fig:HCG88}) shows that HCG88A exhibits a faint plume  visible to the NW of the galaxy. A similar faint extension is not visible on the opposite side of the galaxy, thus the NW feature cannot be explained as a disk warp but it looks similar to tidal tails sometimes seen in interacting galaxies. The length of this faint extension, as the figure demonstrates, is at least 27 kpc and its average surface brightness is $\sigma_R \simeq$ 26 mag/square arcsec. The plume is diffuse and is seen on the deep L-band image combination but not in the H$\alpha$ image (see below) nor in individual images exposed for 200 sec with the CFHT or in individual SDSS images available on NED\footnote{http://ned.ipac.caltech.edu/}. However, the plume is faintly visible on a sum of the $g$, $r$, and $i$ archive images from SDSS.


I note here that the chain of galaxies extending to the north from HCG88B is in the background; these galaxies are at z$\sim$0.19, as demonstrated by their SDSS spectra. The image of HCG88B shows that its outer regions are extended in a direction perpendicular to the elongation of its inner and brighter regions. The shape of the outer regions of this galaxy is visible in the right panel of
Fig.~\ref{fig:HCG88} where the change in position angle of the major axis is seen in the galaxy's isophotes. 
 The contrast-stretched image shown in the left panel shows also similar extensions with abrupt terminations on opposite sides of the galaxy, with the NW boundary closer to the center of the galaxy than the SW boundary. 

Such features, when detected around elliptical galaxies (e.g., Malin \& Carter 1980), have first been interpreted as ``giant shells'' formed by an explosive event in the galaxy followed by shocks in surrounding material.  The presently accepted interpretation is that such shells form when a galaxy merges with another galaxy, by phase-warping a cannibalized disk galaxy via a low angular momentum encounter (Quinn 1984). Such features are not restricted to elliptical galaxies, but can also be found in spirals (e.g., the Umbrella galaxy NGC 4651=Arp 189, e.g., Foster et al. 2014). The stars of the accreted object oscillate in the gravitational potential of the target galaxy while being slowed down by dynamic friction with stars in the target galaxy and the stellar shells form at their turnaround radii. The projectile need not be aligned with the main body of the target galaxy. It is possible that the accreted stars will settle in a configuration oblique to the target galaxy, causing a severe twist to the isophotes. Therefore the features shown in the unsharply-masked image of HCG88B (see Fig.~\ref{fig:HCG88}) could be related to such an event.



HCG88C appears to be an almost face-on two-armed spiral on low-contrast images, but higher dynamic range images show a third, fainter, arm emerging just East of the galaxy center and curling northward. 
 The $\sim$edge-on disk (HCG88D) shows two symmetric ``bumps'' in the L-band intensity image. These could be signs of enhanced star formation at the end of a bar, implying that this object could be a barred spiral.

\begin{figure}[h]
\centering{
  \includegraphics[width=15cm]{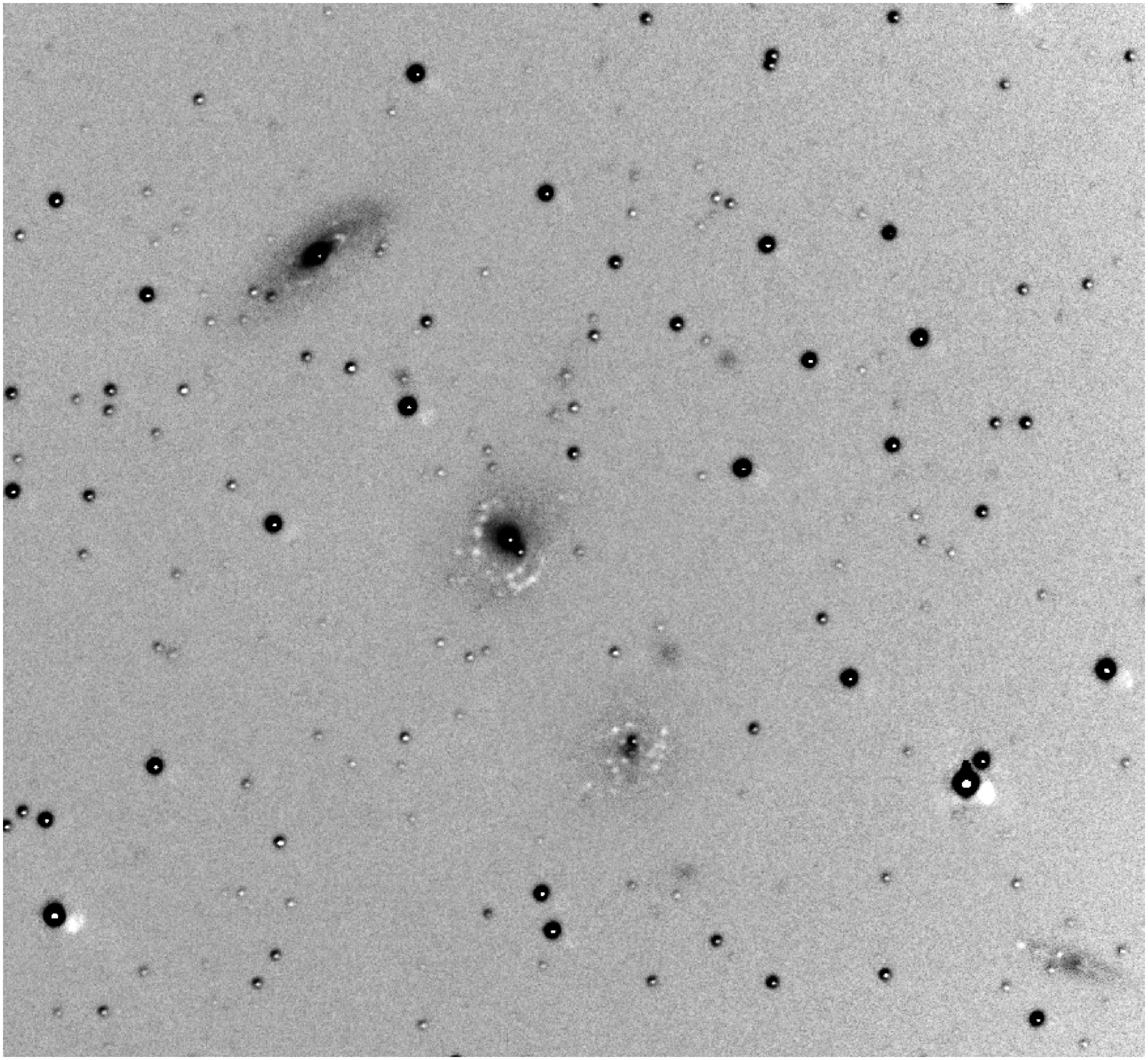}
  \caption{Representation of the entire group emphasizing the HII regions. The ghosts to the lower-right of bright stars, where additional light shows up as white patches, do not affect the conclusion about the HII regions. The HII regions appear as tiny star-like white patches, and should be compared with the broad-band images of the galaxies.}
  \label{fig:HII_regs}}
\end{figure}

Figure~\ref{fig:HII_regs} shows a representation of the HII regions in all four galaxies derived, as explained above, by subtracting a suitably scaled L-band image from the H$\alpha$ image. The PSFs in the two images were deliberately not matched, with the result being that stars appear as dark round patches with a bright centers, whereas HII regions show up only as white patches. This image also shows that the narrow-band H$\alpha$ filter produces faint ghost images adjacent to bright stars; these ghosts are of order 1\% of the intensity of the star and do not influence the conclusion about the distribution of HII regions.

As Figure~\ref{fig:HII_regs} shows, the HII regions are generally confined to the four main group members. None of the nearby galaxies, seen as diffuse patches on the deep L-band image, show H$\alpha$ emission. 
 Neither of the two other immediate members of the group show prominent HII regions.
It is interesting to note how are the HII regions arranged in each object. In HCG88A the distribution of HII regions follows two arcs that are approximately opposite each other, and are relatively close to the central bulge of the galaxy. This matches the H$\alpha$ map of the galaxy shown in Fig. 1a of Plana et al. (2003), where the emission is restricted to the inner 50 arcsec of the galaxy while its major axis is almost two arcmin long.
HCG88B and C show many HII regions. In B these occur mainly to one side of the galaxy, are arranged mostly within one of the spiral arms, and on the outer arm in the same position angle to the SW. This, with the exception of a single HII region visible NW of the center in one of the diffuse patches that 
 corresponds to the outer shell described above. Many of these HII regions can also be recognized in Fig. 2 of Plana et al. (2003). Specifically, the NE arm emission recognized by Plana et al. separates in Figure~\ref{fig:HII_regs} into four bright H$\alpha$ knots and a number of fainter ones. The SW region shows a double row of HII regions starting from the end of the bar and curving to the South. Component D shows a single HII region at its eastern end, closest to the other group galaxies. This is the region at the NE end of the galaxy that is receding in Fig. 5a of Plana et al. (2003). The equatorial dust lane reported by them is also visible in the figure shown here.



\section {Discussion}
\label{sec.discuss}
There are a few indications in the literature that HCG88A may be a post-interaction galaxy.
Chongming \& Tan (1988) analyzed the rotation curve (RC) of HCG88A obtained from Vera Rubin. The rotation curve, shown in their Figure 1, exhibits a distinct asymmetry, with the NW branch having a higher and steeper RC than the SE branch. This asymmetry sets in at $\sim$5 kpc from the center and the RCs for both branches are linear, with the extremal point of the NW branch reaching up to 350 km s$^{-1}$. Chongming \& Tan attributed the RC asymmetry to either the galaxy rotating around a center different from the optical one, or to a perturbation such as a ``jet-like emission'' from the center of the galaxy. However, the observations reported here, which revealed a structure similar to a tidal tail at the NW end of the galaxy, indicate than explanations including an offcenter rotation or a jet from the nucleus can probably be discarded.

The RC of HCG88A and additional data about this galaxy were discussed by Rubin, Hunter \& Ford (1991). The RC is plotted in their Figure 1t and is based on the H$\alpha$ and [NII] emission lines. The galaxy's RC is noted as ``abnormal''; this is emphasized in their Table 4, where the RC asymmetry noted by Chongming \& Tan (1988) is readily visible. A similar asymmetric RC was noted also by Nishimura et al (2000). Rubin, Hunter \& Ford also discuss the optical image characteristics and note that this is asymmetric as well ``with respect to the minor axis, with extended contours to the northwest  in the inner regions, and extended contours to the southwest at lowest light levels''.

Such morphological and kinematical abnormalities could be caused by tidal interactions. Barton, Bromley \& Geller (1999) studied the effects of tidal interactions on rotation curves of galaxies via N-body simulations. Their description, shown graphically in their Figure 1, has a number of asymmetric RVs depending on the viewing direction and encounter parameters. In their Figure 2, where real objects are presented, UGC 484 shows a rising RV in the direction of a faint extension to the galaxy, similar to that of HCG88A. That, however, could be associated with the bar in this object. The authors note that such kinematic features do not require dissipative matter in order to form. Note also that the kinematic anomaly would not have been revealed by the imaging Fabry-P\'{e}rot studies of the H$\alpha$ line (e.g. Amram et al. 2002), since for HCG88A the faint feature is seen in the continuum, not in the emission line image.

While the presence of a plume in HCG88A indicates that some kind of interaction did take place, it would be interesting to be able to identify which other object was involved in this interaction and when did this process happen. For this, I turn to simulations of galaxy interactions, specifically to those of Oh, Kim and Lee (2015). Theirs is a self-consistent 3D N-body simulation using GADGET, where the objects are a Milky-Way like disk galaxy with a bulge and a halo, which interacts with a smaller mass halo. I compare the appearance of HCG88A on my images with those shown in their Figure 2, which represent different interaction stages. Given that HCG88A shows no counter-tail, and neither can the putative interacting companion be easily distinguished in the tail, a possible match could be for a time step t$\simeq$0.1 (i.e., about 100 Myr following the closest approach) when the tail is relatively bright, with the interactor hiding in the plume.

Verdes-Montenegro et al. (2001) mapped a number of HCGs in HI. HCG88 was observed with the VLA in configuration C. They found a total HI mass for the group of $1.9\times10^{10}$ M$_{\odot}$ whereas the predicted total HI was slightly higher at 3.5$\times10^{10}$ M$_{\odot}$, implying a slight HI deficiency. The HI deficiencies of components A and B were higher than those of C and D, with possibly 2.9$\times10^{8}$ M$_{\odot}$ of HI claimed to be in tidal features external to the galaxies. The detailed HI map of the group, shown at the top of their Fig. 7, indicates that the HI in HCG88A is concentrated near the center of the galaxy, approximately where the arcs of HII regions are detected (see Figure~\ref{fig:HII_regs}). HCG88B and C show connected HI contours in their outer regions. While HCG88B shows approximately the same HI distribution as that of the optical light, C has a more extended HI distribution, about twice that of the optical image. 
 Component D shows an HI tail pointing away from the group and its HI contours in the direction of the rest of the group appear more compressed than in the opposite direction.

Borthakur, Yun \& Verdes-Montenegro (2010) observed HCG88 with the GBT telescope in the HI and CO lines. The $\sim$9.1 arcmin synthesized beam of this telescope is much larger than the extent of the four main galaxies of the group thus this observation has the potential of measuring diffuse flux missed by the VLA interferometer with its 24$\times$17 arcsec$^2$ beam. They measured a total HI mass of 11.9$\pm 0.2 \times10^{9}$ M$_{\odot}$, similar to the value measured with the VLA, implying that the suspicion of a more extended HI distribution cannot be confirmed.

Two galaxies, HCG88A and C, were detected by IRAS at 60 and 100 $\mu$m (Allam et al. 1996): A[60]=0.44 Jy and A[100]=2.73 Jy, and C[60]=0.32 Jy and C[100]=2.23 Jy respectively. All galaxies are visible on WISE IR images, in all four bands, but the data are too shallow and of low-resolution to reveal the tidal tail of HCG88A or details in the disks. However, the faint features of Component B seen in the L-band image and extending perpendicular to the main body of the galaxy are also visible in unsharply-masked W1 (3.4 $\mu$m) and W2 (4.6 $\mu$m) images.

Since the observed velocity dispersion of HCG88 is exceedingly low, and assuming that the transversal velocity dispersion is similar to that in the line-of-sight, it follows that the crossing time (which is of order the dynamical time for the group) is about one Hubble time. Thus the interactions apparently seen in the galaxies of this group were not between the main visible objects, but presumably with smaller objects that are not detected at present. In particular, HCG88B which shows shells and weird isophotes probably accreted the object it interacted with at least 1 Gyr ago (a few crossing times of this galaxy) and during this period the oscillations of the accreted object formed the shells and isophote extensions. The remnant could hide now in the northern part of this galaxy. I note that while the interaction signatures are somewhat clear for HCG88A and HCG88B, they are less visible or not visible at all in the two other main galaxies of the group.

Finally, it is argued that interactions may enhance the activity of galaxy nuclei (e.g., Kouolouridis et al. 2013). Coziol et al. (1998) studied the nuclear activity of the brightest members of Hickson Compact Groups using spectroscopy and found that HCG88A is a dwarf Seyfert 2, while HCG88B is a dwarf liner, as per their Table 2. Both galaxies are low-luminosity AGNs, with the full-width at half-maximum of the H$\alpha$ line being 704 km s$^{-1}$ and 517 km s$^{-1}$ respectively. 
 Shimada et al. (2000) confirmed that Component B is an AGN, based on the strength of the [NII]/H$\alpha$ ratio.

Menon (1995) mapped HCG88 with the VLA in the radio continuum at 1.4 GHz and found that only component A is a source, with a total flux density of 0.90 mJy, all coming from the central regions of the galaxy. Leon, Combes and Menon (1998) detected HCG88A in CO with the IRAM 30m radiotelescope on Pico Veleta, deriving a total H$_2$ mass of 6.2$\times10^9$ M$_{\odot}$. Other galaxies in  HCG88 were not targeted in this single-beam CO survey.

Martinez-Badenes et al. (2012) studied the molecular gas and star formation properties in HCGs via CO observations with IRAM or from the literature. They re-extracted the IRAS FIR photometry by co-adding calibrated scans to reach brightness levels a few times deeper than the IRAS Point Source Catalog. For HCG88 they list results only for the four principal galaxies; for component C they find a significant excess of observed vs. predicted quantities such as H$_2$, L$_{FIR}$ and M(HI). They note ``that half of [the objects with strong deficiencies] present strong signs of distortion (tidal tails in the optical and/or HI, kinematical perturbations, etc.)''. The star formation rates they find are of order a few M$_{\odot}$ yr$^{-1}$, with the exception of HCG88B for which SFR$\leq0.77$ M$_{\odot}$ yr$^{-1}$.

Despite these already-published hints that HCG88 witnessed some galaxy interactions, there have been claims in the literature that such interactions did not happen. Mamon (2000) noted that the HCG88 group has an anomalously low velocity dispersion. This too low a velocity dispersion is, according to Mamon, a sign that  the group is not a dense system near virialization, but rather a chance alignment of galaxies in a loose group near turnaround, when the velocity dispersion is expected to be small, or a case where the tidal friction slowed down the galaxies considerably.


The results presented here, specifically the faint surface brightness extension of HCG88A that is not associated with HI, the chaotic arms of HCG88B, the position angle shift between the inner and the outer isophotes, and the shell-like structure of this galaxy, and possibly also the faint third arm of HCG88C, contradict the claims of Verdes-Montenegro et al. (2001) and of Da Rocha \& Mendes-de Olivera (2005) that the group is at an early stage of interaction, or is not interacting at all. 
Contrary to the proposal of Mamon (2000a), the observations reported here do show that the group is probably real, the galaxies are or have been interacting, although not with the group members now visible, and N6977 may also have accreted a smaller galaxy in an event that possibly produced both the outer shells as well as activating a distant HII region on its northern part.

The findings presented here do not support the proposal of Plauchu-Frayn \& Coziol (2007) that in a group, the galaxies first lose their gas through galaxy-galaxy interactions and only afterwards they start merging under ``dry'' conditions. The galaxy interactions might also have caused the activation of the liner in N6977 and the Seyfert 2 in N6978, as well as producing the tidal tail.
The object that interacted with HCG88A, and the one that was probably accreted by HCG88B, could represent an evolutionary stage when galaxies are built up by the inclusion of small dwarf galaxy sized entities.


\section{Summary}
\label{sec.summary}
I presented results from deep broad-band and rest-frame H$\alpha$ imaging of the Hickson Compact Group 88 from which I conclude that the group is real and has suffered significant interactions but not with the present group members, at least for the two objects at its northern half. This first paper presenting results from the JBRT telescope emphasizes its capabilities in low surface brightness imaging.

\section{Acknowledgements}
I am grateful to Mike Rich, Mike Shara, and Ezra Drucker for their financial contributions that made the acquisition of the JBRT possible. The construction of the JBRT and its dome at the Wise Observatory was possible through the hard work of the WO staff members Shai Kaspi, Ezra Mash'al and Sammy Ben Guigui, as well as of the ``WO friends'' Ilan Manulis, Assaf Berwald, and former graduate students Dr. Evgeny Gorbikov and Dr. Ido Finkelman. I am also grateful to Dr. Mischa Schirmer for his hand-in-hand help in implementing and running THELIGUI at the Wise Observatory.

Funding for the SDSS and SDSS-II has been provided by the Alfred P. Sloan Foundation, the Participating Institutions, the National Science Foundation, the U.S. Department of Energy, the National Aeronautics and Space Administration, the Japanese Monbukagakusho, the Max Planck Society, and the Higher Education Funding Council for England. The SDSS Web Site is http://www.sdss.org/.

The SDSS is managed by the Astrophysical Research Consortium for the Participating Institutions. The Participating Institutions are the American Museum of Natural History, Astrophysical Institute Potsdam, University of Basel, University of Cambridge, Case Western Reserve University, University of Chicago, Drexel University, Fermilab, the Institute for Advanced Study, the Japan Participation Group, Johns Hopkins University, the Joint Institute for Nuclear Astrophysics, the Kavli Institute for Particle Astrophysics and Cosmology, the Korean Scientist Group, the Chinese Academy of Sciences (LAMOST), Los Alamos National Laboratory, the Max-Planck-Institute for Astronomy (MPIA), the Max-Planck-Institute for Astrophysics (MPA), New Mexico State University, Ohio State University, University of Pittsburgh, University of Portsmouth, Princeton University, the United States Naval Observatory, and the University of Washington.

Remarks by an anonymous referee improved significantly the  presentation, and are acknowledged.

\begin{description}{}

\item Allam S., Assendorp R., Longo G., Braun M., Richter G., 1996, A\&AS, 117, 39

\item Amram P., Mendes de Oliveira C., Plana H., Balkowski C., Boulesteix J., Carignan C., 2002, Astrophys. Space Sci., 281, 389

\item Barton E.~J., Bromley B.~C., Geller M.~J., 1999, ApJ, 511, L25

\item Borthakur S., Yun M.~S., Verdes-Montenegro L., 2010, ApJ, 710, 385

\item Brosch, N., Kaspi, S., Niv, S., Manulis, I., 2015, Astrophys. Space Sci., in press

\bibitem[\protect\citeauthoryear{Chongming
\& Tan}{1988}]{1988Ap&SS.145...47C} Chongming X., Tan L., 1988, Astrophys. Space Sci., 145, 47

\item Cooper A.~P., et al., 2010, MNRAS, 406, 744

\item Coziol R., Ribeiro A.~L.~B., de Carvalho R.~R., Capelato H.~V., 1998, ApJ,
493, 563

\item Da Rocha C., Mendes de Oliveira C., 2005, MNRAS, 364, 1069

\item de Carvalho R.~R., Ribeiro A.~L.~B.,
Capelato H.~V., Zepf S.~E., 1997, ApJS, 110, 1

\item D{\'{\i}}az-Gim{\'e}nez E., Mamon G.~A., Pacheco M., Mendes de Oliveira C., Alonso M.~V., 2012, MNRAS, 426, 296

\item Erben T., et al., 2005, Astron. Nachrichten, 326, 432

\item Fakhouri O., Ma C.-P., Boylan-Kolchin M., 2010, MNRAS, 406, 2267

\item Foster C., et al., 2014, MNRAS, 442, 3544

\item Hernquist L., Katz N., Weinberg D.~H., 1995, ApJ, 442, 57

\item Hickson P., 1982, ApJ, 255, 382  (erratum ApJ, 259, 930)

\item Hickson P., Mendes de Oliveira C., Huchra J.~P., Palumbo G.~G., 1992, ApJ, 399, 353

\item Ibata R.~A., et al., 2013, Natur, 493, 62

\item Koulouridis E., Plionis M., Chavushyan V., Dultzin D., Krongold Y., Georgantopoulos I., Le{\'o}n-Tavares J., 2013, A\&A, 552, A135

\item Leon S., Combes F., Menon T.~K., 1998, A\&A, 330, 37

\item Malin D.~F., Carter D., 1980, Nature, 285, 643

\item Martin N.~F., Ibata R.~A., McConnachie A.~W., Mackey A.~D., Ferguson
A.~M.~N., Irwin M.~J., Lewis G.~F., Fardal M.~A., 2013, ApJ, 776, 80

    \item Mamon
G.~A., 2000, {\it Dynamics of Galaxies: from the Early Universe to the present}, F. Combes, G.A. Mamoon \& V. Charmandaris, eds., ASP Conference Series 197, 377 

\item Martinez-Badenes V., Lisenfeld U., Espada D., Verdes-Montenegro L., Garc{\'{\i}}a-Burillo S., Leon S., Sulentic J., Yun M.~S., 2012, A\&A, 540, A96

\item McQuinn K.~B.~W., et al., 2014, ApJ, 785, 3

\item Menon
T.~K., 1995, MNRAS, 274, 845

\item Nishiura S., Shimada M., Ohyama Y.,
Murayama T., Taniguchi Y., 2000, AJ, 120, 1691

\item Oh S.~H., Kim W.-T., Lee H.~M., 2015, ApJ, 807, 73

\item Plana H., Amram P., Mendes de Oliveira C., Balkowski C., Boulesteix J.,
2003, AJ, 125, 1736

\item Plauchu-Frayn I., Coziol R., 2008, {\it Formation and Evolutioin of Galaxy Bulges} IAU Symposium 245, M. Bureau, E. Athanassoula \& B. Barbuy eds.,  p85

\item Purcell C.~W., Kazantzidis S., Bullock J.~S., 2009, ASPC, 419, 248

\item Purcell C.~W., Bullock J.~S., Kazantzidis S., 2010, MNRAS, 404, 1711

\item Quinn
P.~J., 1984, ApJ, 279, 596

\item Ribeiro A.~L.~B., de Carvalho R.~R.,
Capelato H.~V., Zepf S.~E., 1998, ApJ, 497, 72

\item Rich R.~M., Collins M.~L.~M., Black C.~M., Longstaff F.~A., Koch A., Benson
A., Reitzel D.~B., 2012, Nature, 482, 192

\item Rubin V.~C., Hunter D.~A., Ford W.~K., Jr., 1991, ApJS, 76, 153


\item Schirmer M., 2013, ApJS, 209, 21

\item Shimada M., Ohyama Y., Nishiura S., Murayama T., Taniguchi Y., 2000, AJ, 119, 2664

\item Spector O., Finkelman I., Brosch N., 2012, MNRAS, 419, 2156

\item Torres-Flores S., Mendes de Oliveira C., Plana H., Amram P., Epinat B., 2013, MNRAS, 432, 3085

\item Zitrin A., Brosch N., 2008, MNRAS, 390, 408

\item Verdes-Montenegro L., Yun M.~S., Williams B.~A., Huchtmeier W.~K., Del Olmo A., Perea J., 2001, A\&A, 377, 812

\item Watkins A.~E., Mihos J.~C., Harding P., 2015, ApJ, 800, LL3

\end{description}

\end{document}